\documentclass[12pt]{article}
\usepackage{a4wide}
\usepackage{latexsym}

\usepackage{pslatex}
\usepackage[latin1]{inputenc}
\usepackage[T1]{fontenc}
\usepackage{axodraw}
\usepackage{epsfig}

\def\bq{\begin{eqnarray}}
\def\eq{\end{eqnarray}}

\def\eps{\varepsilon}
\def\v{\verb}
\def\gversion{1.0.0 }

\newlength{\dinwidth} \newlength{\dinmargin}
\begin{document}
\thispagestyle{empty}

\begin{flushright}
  UPRF-2003-06\\
  IHES/P/03/24
\end{flushright}

\vspace{1.5cm}

\begin{center}
  {\Large\bf gTybalt - a free computer algebra system\\}
  \vspace{1cm}
  {\large Stefan Weinzierl}\\
  \vspace{1cm}
  {\small {\em Dipartimento di Fisica, Universit\`a di Parma,\\
       INFN Gruppo Collegato di Parma, 43100 Parma, Italy}\\
                        and \\
          {\em Institut des Hautes Etudes Scientifique,
               91440 Bures-sur-Yvette, France}
  } \\
\end{center}

\vspace{2cm}

\begin{abstract}\noindent
  {%
    This article documents the free computer algebra system ``gTybalt''.
    The program is build on top of other packages, among others
    GiNaC, TeXmacs and Root.
    It offers the possibility of interactive symbolic calculations
    within the C++ programming language.
    Mathematical formulae are visualized using TeX fonts.
   }
\end{abstract}

\vspace*{\fill}

\newpage 

{\bf\large PROGRAM SUMMARY}
\vspace{4mm}
\begin{sloppypar}
\noindent   {\em Title of program\/}: gTybalt \\[2mm]
   {\em Version\/}: \gversion \\[2mm]
   {\em Catalogue number\/}: \\[2mm]
   {\em Program obtained from\/}: {\tt http://www.fis.unipr.it/\~{}stefanw/gtybalt} \\[2mm]
   {\em E-mail\/}: {\tt stefanw@fis.unipr.it} \\[2mm]
   {\em License\/}: GNU Public License \\[2mm]
   {\em Computers\/}: all \\[2mm]
   {\em Operating system\/}: GNU/Linux \\[2mm]
   {\em Program language\/}: {\tt C++     } \\[2mm]
   {\em Memory required to execute\/}: 
         64 MB recommended \\[2mm]
   {\em Other programs called\/}: see appendix \ref{sec:installation} \\[2mm]
   {\em External files needed\/}: none \\[2mm]
   {\em Nature of the physical problem\/}: 
         Symbolic calculations occur nowadays in all areas of science.
         gTybalt is a free computer algebra system based on the C++ language.\\[2mm]
   {\em Method of solution\/}: 
         gTybalt is a ``bazaar''-style program, it relies on exisiting, 
         freely-available packages
         for specific sub-tasks.\\[2mm] 
   {\em Restrictions on complexity of the problem\/}: 
         gTybalt does not try to cover every domain of mathematics.
         Some desirable algorithms, like symbolic integration are not 
         implemented.
         It can however easily be extended in new directions.
         Apart from that, standard restrictions due to the 
         available hardware apply. \\[2mm]
   {\em Typical running time\/}:
         Depending on the complexitiy of the problem.
\end{sloppypar}

\newpage

\reversemarginpar

{\bf\large LONG WRITE-UP}

\section{Introduction}
\label{sec:intro}

Symbolic calculations, carried out by computer algebra systems,
have become an integral part in the daily work of scientists.
The advance in algorithms and 
computer technology has led to remarkable progress 
in several areas of natural sciences.
A striking example is provided by the tremendous progress 
in the last few years for analytic calculations of
so-called loop diagrams in perturbative quantum field theory.
It is worth to analyse what the particular requirements on a computer
algebra systems for these calculations are:
First of all, these tend to be ``long'' calculations, e.g. the 
system needs to process large amounts of data and efficiency in performance
is a priority.
Secondly, the algorithms for the solution of the problem are usually
developed and implemented by the physicists themselves.
This requires support from the computer algebra system for a programming
language which allows to implement complex algorithms for abstract
mathematical entities.
In other words, it requires support of object oriented programming
techniques from the system.
On the other hand, these calculations usually do not require
that the computer algebra system provides sophisticated tools
for all branches of mathematics.
Thirdly, despite the fact that these calculations process large amounts
of data, the time needed for the implementation of the 
algorithms usually outweights the actual running time of the program.
Therefore convenient development tools are also important.
Here I report on the program ``gTybalt'', a free computer algebra
system.
The main features of gTybalt are:
\begin{itemize}
\item Object Oriented: gTybalt allows symbolic calculations within
the C++ programming language.
\item Efficiency for large scale problems: Solutions developed with gTybalt can
be compiled with a C++ compiler and executed independently of gTybalt.
This is particular important for computer-extensive problems and a major 
weakness of commercial computer algebra systems.
\item Short development cycle: gTybalt can interpret C++ and execute C++ scripts.
Solutions can be developed quickly for small-scale problems, either interactively 
or through scripts, and once
debugged, the solutions can be compiled and scaled up to large-scale problems.
\item High quality output: Mathematical formulae are visualized using TeX fonts
and can easily be converted to LaTeX on a what-you-see-is-what-you-get basis.
\end{itemize}
gTybalt is a free computer algebra system and distributed under the terms
and conditions of the GNU General Public Licence.
Compared to other computer algebra systems, it does not
try to cover every domain
of mathematics.
Some desirable algorithms, like symbolic integration are not implemented.
However, the modular design of gTybalt allows to incorporate easily new
algorithms.

The functionality of a computer algebra system can be divided into different
modules, e.g. there will be a module, which displays the output, a second module
analysis and interprets the input, a third module does the actual symbolic
calculation.
Writing a computer algebra system from scratch is a formidable task.
Fortunately it is not required, since there are already freely available
packages for specific tasks.
gTybalt is based on several other packages and provides the necessary communication
mechanisms among these packages.
gTybalt is therefore a prototype of a ``bazaar''-style program 
\cite{Raymond} and an example
of what is possible within the free software community.
It should be clearly stated, that without these already existing packages
gTybalt would never have been developed and my thanks go to the authors
of these packages for sharing their programs with others.
In particular gTybalt is build on the following packages:
\begin{itemize}
\item The TeXmacs-editor 
\cite{vanderHoeven:2001}
is used to display the output
of formulae in high quality mathematical typesetting using TeX fonts.
\item gTybalt can also be run from a text window. Then the library
eqascii 
\cite{eqascii}
is used to render formulae readable in text mode.
\item Any interactive program needs an interpreter for its commands.
gTybalt uses the CINT C/C++ interpreter
\cite{Goto}, 
which allows execution of C++ scripts and C++ command line input.
\item At the core of any computer algebra system is the module for
symbolic and algebraic manipulations. This functionality is provided
by the GiNaC-library
\cite{Bauer:2000cp}.
\item One aspect of computer algebra systems is arbitrary precision arithmetic.
Here GiNaC (and therefore gTybalt) relies on the Class Library for Numbers
(CLN library)
\cite{CLN}.
\item Plotting functions is very helpful to quickly visualize results.
The graphic abilities of gTybalt are due to the
Root-package
\cite{Brun:1996}.
\item The GNU scientific library is used for Monte Carlo integration
\cite{GSL}.
\item Optionally gTybalt can be compiled with support for the expansion of transcendental
 functions.
This requires the nestedsums library 
\cite{Weinzierl:2002hv}
to be installed.
\item Optionally gTybalt can be compiled with support for factorization of polynomials.
This requires the NTL library 
\cite{NTL}
to be installed.
\end{itemize}
Additional documentation:
The manual of gTybalt, which comes with the distribution,
provides additional information for this program.
Since gTybalt incoporates GiNaC, Root and TeXmacs,
you are advised to read also the documentation on GiNaC, Root and TeXmacs.
A general introduction to computer algebra can be found in 
\cite{Weinzierl:2002cg}
and in the references therein.

This article is organized as follows:
The next section gives a brief introduction on how to use the program.
Sect. \ref{sec:examples} demonstrates for some simple examples how
gTybalt can be used for calculations in particle physics.
This is the only section which requires some background in particle physics
and readers not familiar with this topic may skip this section.
Sect. \ref{sec:design} gives details on the design of the program and serves
as a guide to the source code.
Finally, sect. \ref{sec:conclusions} summarizes this article.
In an appendix I give some detailed hints for the installation of the program. 

\section{Tutorial for gTybalt}
\label{sec:running}

This section gives a short tutorial introduction to gTybalt by discussing
some small and simple examples.

\subsection{Starting gTybalt}

gTybalt can be run in the TeXmacs mode or in a simple text mode.
To start gTybalt in text mode, type 
\v/gtybalt/.
To quit, type 
\v/quit/.
To use gTybalt within TeXmacs, first start TeXmacs with the command
\v/texmacs/.
You can then start a gTybalt session by clicking on the terminal symbol
and selecting ``gTybalt'' from the pop-up menu.
Alternatively you can start gTybalt from the ``Text'' menu via
``Text $\rightarrow$ Session $\rightarrow$ GTybalt''.

\subsection{Command line input}

\begin{figure}
\centerline{
\epsfig{file=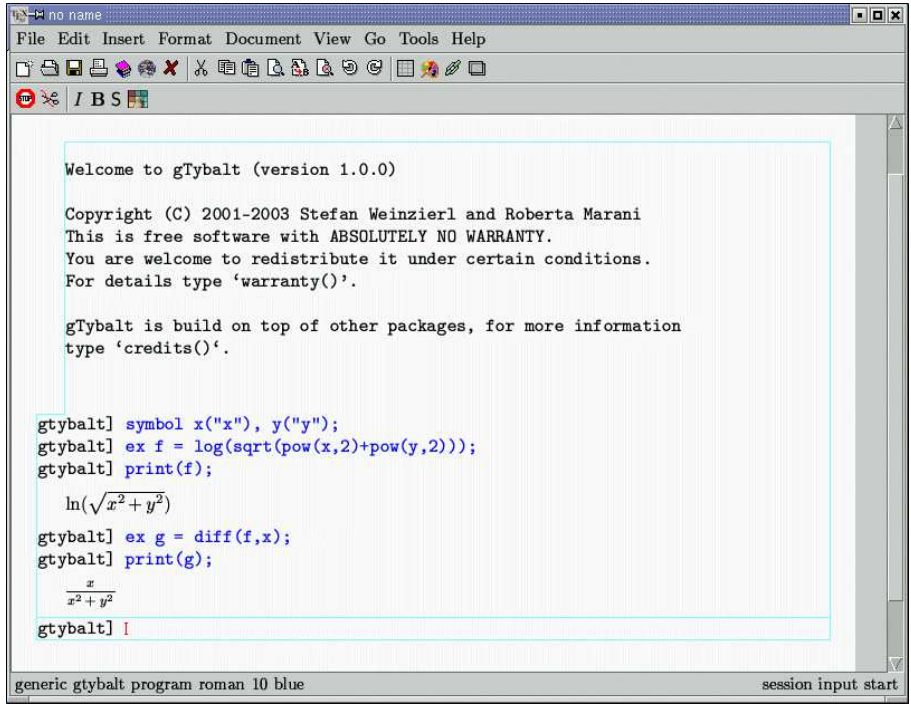, width=12cm}
}
\caption{\label{fig1} A screenshot for gTybalt when running in TeXmacs mode.}
\end{figure}

\begin{figure}
\centerline{
\epsfig{file=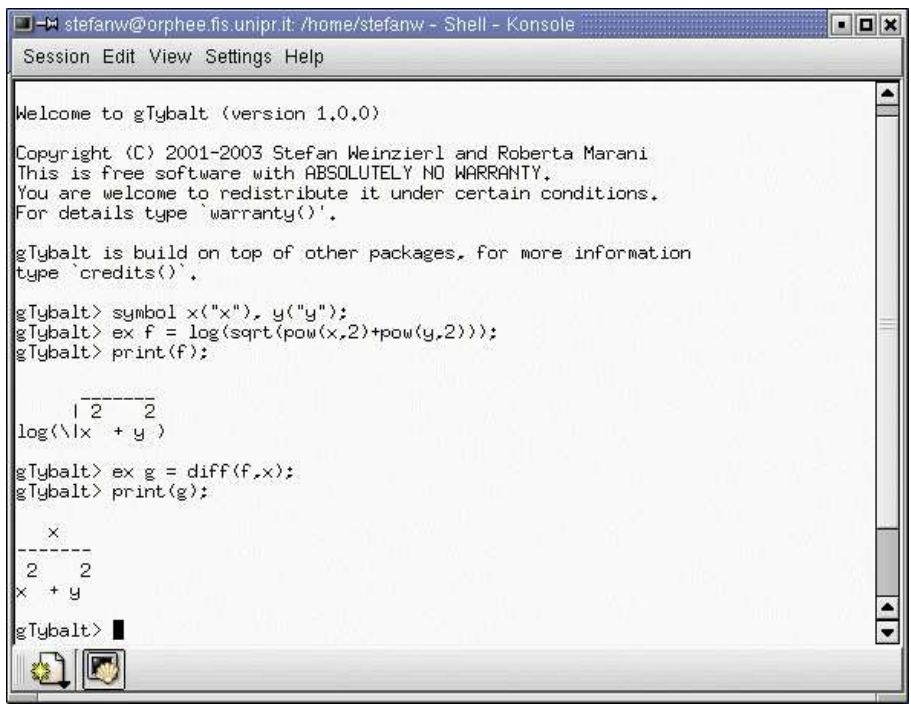, width=12cm}
}
\caption{\label{fig2} A screenshot for gTybalt when running in text mode.}
\end{figure}

You can type in regular C++ statements which will be processed by
CINT. For example
\begin{verbatim}
gTybalt> int i=1;
gTybalt> i++;
gTybalt> cout << "The increased number : " << i << endl;
The increased number : 2
\end{verbatim}
The functionality of gTybalt for symbolic and algebraic calculations 
is provided by the GiNaC-library.
The syntax follows the one for the GiNaC-library.
For example:
\begin{verbatim}
gTybalt> symbol a("a"), b("b");
gTybalt> ex e1=pow(a+b,2);
gTybalt> print(e1);

     2
(b+a)

gTybalt> ex e2=expand(e1);
gTybalt> print(e2);

 2    2
a  + b  + 2 a b

\end{verbatim}
Here \v/print/ is a gTybalt-subroutine, which prints a variable to the screen.
By default, gTybalt does not print anything onto the screen, unless
the user specifically asks for a variable to be printed.
If gTybalt is running under TeXmacs, the output will be with TeX fonts.
There is also a function \v/rawprint/ which prints the variable
\v/e2/ as follows:
\begin{verbatim}
gTybalt> rawprint(e2);
a^2+b^2+2*a*b 
\end{verbatim}
Fig. \ref{fig1} shows how the output of a further example will 
look like under TeXmacs.
Fig. \ref{fig2} shows the corresponding output, when gTybalt runs in text mode.
Within TeXmacs mode there is the possibility to print a session to a 
postscript file by choosing from the ``File'' menu the combination
``File $\rightarrow$ Export $\rightarrow$ Postscript''.
It is also possible to generate for a session a corresponding
LaTeX file via
``File $\rightarrow$ Export $\rightarrow$ LaTeX''.
This is in particular useful if one would like to obtain
for a displayed formula the corresponding LaTeX code.

\subsection{Scripts}

The standard behaviour of the C++/C interpreter CINT is to interpret
any command immediately.
There is also the possibility to put a few commands into a script
and to load this file into a session.
This is done through the following commandds:
\begin{verbatim}
.L file.C
.x file.C
\end{verbatim}
The \v/.L/ command loads a script into the session, but does not execute
the script. This is useful for a script containing the definition of
a function.
The \v/.x/ command loads and executes a script.
As an example consider that the file
\v/hermite.C/ contains the following code:
\begin{verbatim}
ex HermitePoly(const symbol & x, int n)
{
 ex HKer=exp(-pow(x,2));
 return normal(pow(numeric(-1),n) * diff(HKer,x,n)/HKer);
}
\end{verbatim}
This is just a function which calculates the $n$-th Hermite polynomial.
Now try the following lines in gTybalt:
\begin{verbatim}
gTybalt> .L hermite.C
gTybalt> symbol z("z");
gTybalt> ex e1=HermitePoly(z,3);
gTybalt> print(e1);

             3
 - 12 z + 8 z

\end{verbatim}
This prints out the third Hermite polynomial.
As a further example let the file 
\v/main.C/ contain the following code:
\begin{verbatim}
{
 symbol z("z");
 for (int i=0;i<5;i++)
  {
   print( HermitePoly(z,i));
  }
}
\end{verbatim}
This is called an ``un-named script''. Un-named scripts 
have to start with an opening ``$\{$'' and end with a closing ``$\}$''.
Then the following lines in gTybalt
\begin{verbatim}
gTybalt> .L hermite.C
gTybalt> .x main.C
\end{verbatim}
will print the first five Hermite polynomials.

\subsection{Plots}

A function can be plotted as follows:
\begin{verbatim}
gTybalt> symbol x("x");
gTybalt> ex f1=sin(x);
gTybalt> plot(f1,x,0,20); 
\end{verbatim}
This will plot 
$\sin(x)$ 
in the intervall from 0 to 20.
\begin{figure}
\centerline{
\epsfig{file=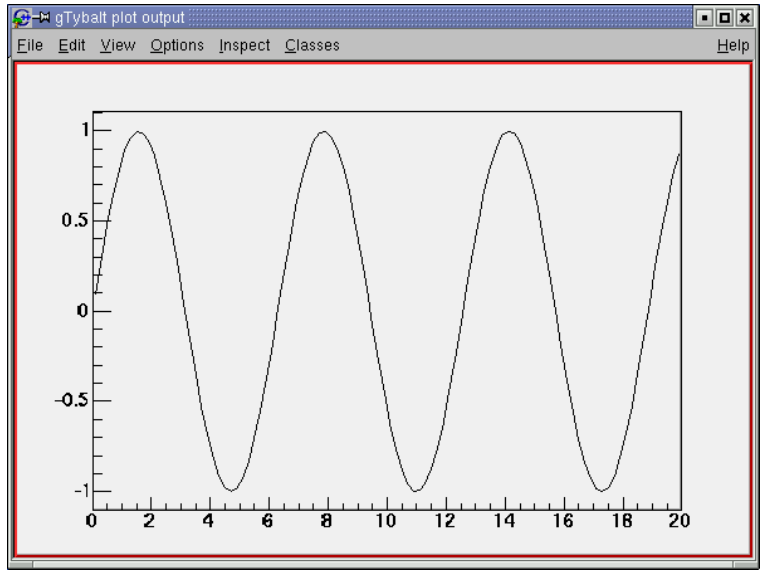, width=10cm}
}
\caption{\label{fig3} The plot for the function $\sin(x)$ for $x$ from $0$ to $20$.}
\end{figure}
To clear the window with the plot, choose from the menubar
of the plot 
``File $\rightarrow$ Quit ROOT''.
\\
Similar, a scalar function of two variables can be plotted as follows:
\begin{verbatim}
gTybalt> symbol x("x"), y("y");
gTybalt> ex f2=sin(x)*sin(y);
gTybalt> plot(f2,x,y,0,10,0,20); 
\end{verbatim}
This will plot 
$\sin(x) \sin(y)$ for $x$ from $0$ to $10$ and $y$ from $0$ to $20$.
Fig. \ref{fig4} shows the output from the plotting routine.
\begin{figure}
\centerline{
\epsfig{file=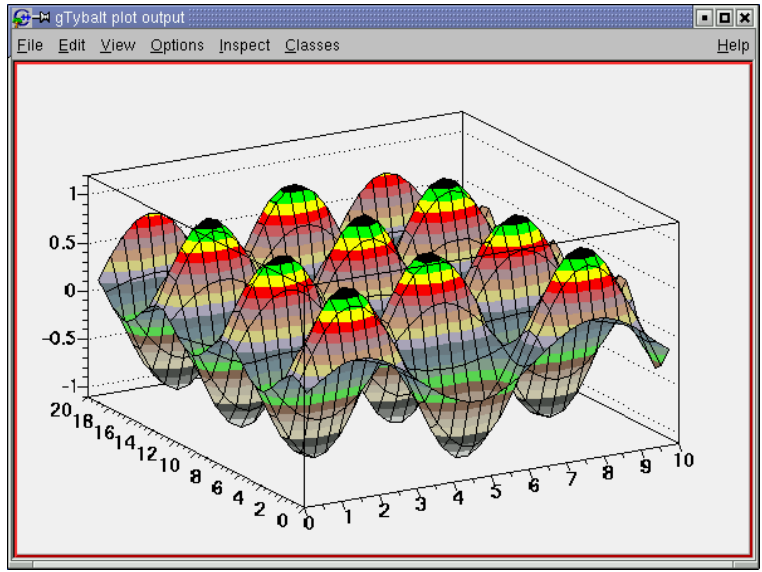, width=10cm}
}
\caption{\label{fig4} The plot for the function $\sin(x) \sin(y)$ for $x$ from $0$ to $10$ and $y$ from $0$ to $20$.}
\end{figure}
To view the plot from a different angle, just grap the plot with the mouse
and move it around.
There is a wide variety of options on how to draw a graph.
To access the draw panel, click on the right mouse button, when the mouse
is placed inside the window containing the plot and choose ``DrawPanel''
from the pop-up menu.
The options include among others lego- and contour-plots. 
The default option corresponds to the style ``surf'' and draws a (coloured)
surface.
\\
The plot can be saved to a file. For example, to save the plot
as a postscript file, choose from the ``File'' menu the option
``Save As canvas.ps''.

\subsection{Numerical integration}

Functions can be integrated numerically by Monte Carlo integration.
For example to evaluate the integral
\bq
\int\limits_0^1 dx 
\int\limits_0^1 dy
\int\limits_0^1 dz
\; x y z 
\eq
one types
\begin{verbatim}
gTybalt> symbol x("x"), y("y"), z("z");
gTybalt> ex f = x*y*z;
gTybalt> ex g = intnum(f,lst(x,y,z),lst(0,0,0),lst(1,1,1)); 
gTybalt> print(g);
0.12500320720479463077
\end{verbatim}
The result of the integration can also be accessed with the help
of the global variable 
\\
\v/gTybalt_int_res/.
In addition the global
variables \v/gTybalt_int_err/ and 
\\
\v/gTybalt_int_chi2/
give information on the error and the 
$\chi^2$.
For our example, one gets
\begin{verbatim}
gTybalt> print(gTybalt_int_res);
0.125003
gTybalt> print(gTybalt_int_err);
5.26234e-06
gTybalt> print(gTybalt_int_chi2);
0.385639
\end{verbatim}
The Monte Carlo integration uses the adaptive
algorithm VEGAS \cite{Lepage:1978sw}.
gTybalt uses the implementation from the GNU Scientific Library.
The algorithm first uses \v/gTybalt_int_iter_low/ iterations
with \v/gTybalt_int_calls_low/ calls to obtain some rough information
where the integrand is largest in magnitude.
The results of this run are discarded, but the grid is kept.
The algorithm then performs \v/gTybalt_int_iter_high/ iterations
with \v/gTybalt_int_calls_high/ calls to obtain a Monte Carlo
estimate.
The default values are
\begin{verbatim}
gTybalt> print(gTybalt_int_iter_low);
5
gTybalt> print(gTybalt_int_calls_low);
10000
gTybalt> print(gTybalt_int_iter_high);
10
gTybalt> print(gTybalt_int_calls_low);
100000
\end{verbatim}
The values of these variables can be adjusted by the user.
For functions of one or two variables there are in addition 
the following simpler forms
\begin{verbatim}
double intnum(ex expr, ex x, double xmin, double xmax);
double intnum(ex expr, ex x, ex y, double xmin, double xmax,
                                   double ymin, double ymax);
\end{verbatim}

\subsection{Factorization}

When gTybalt is compiled with the NTL library, gTybalt
provides an interface to factorize 
univariate polynomials with integer coefficients.
For example
\begin{verbatim}
gTybalt> symbol x("x");
gTybalt> ex f = expand( pow(x+2,13)*pow(x+3,5)*pow(x+5,7)*pow(x+7,2) );
gTybalt> ex g = factorpoly(f,x);
gTybalt> print(g);

        2        5        7        13
 (7 + x)  (3 + x)  (5 + x)  (2 + x)

\end{verbatim}
If the first argument of the function \v/factorpoly/ is not an 
univariate polynomial with integer coefficients, it returns
unevaluated.

\subsection{Expansion of transcendental functions}

When gTybalt is compiled with the nestedsums library, gTybalt
provides an interface to expand a certain class of transcendental
functions in a small parameter.
The class of functions comprises among others generalized hypergeometric functions,
the first and second Appell function and the first Kampe de Feriet function. 
A hypergeometric function can be expanded as follows:
\begin{verbatim}
gTybalt> symbol x("x"), eps("epsilon");
gTybalt> transcendental_fct_type_A F21(x,lst(1,-eps),lst(1-eps),lst(1-eps),
         lst(1,-eps));
gTybalt> ex f = F21.set_expansion(eps,5);
gTybalt> rawprint(f);
-Li(3,x)*epsilon^3-Li(2,x)*epsilon^2-Li(4,x)*epsilon^4-Li(1,x)*epsilon
+Z(Infinity)
\end{verbatim}
This expands the hypergeometric function
${}_2F_1(1,-\varepsilon;1-\varepsilon;x)$ in $\varepsilon$
up to order 5 and agrees with the known expansion
\bq
{}_2F_1(1,-\varepsilon;1-\varepsilon;x)
 & = & 
 1 
 - \eps \;\mbox{Li}_1(x)
 - \eps^2 \;\mbox{Li}_2(x)
 - \eps^3 \;\mbox{Li}_3(x)
 - \eps^4 \;\mbox{Li}_4(x)
 + {\cal O}(\eps^5).
\eq
The algorithm for the expansion is based on an algebra for nested sums
\cite{Moch:2001zr}.
Z(Infinity) represent the unit element in this algebra and is equal to 1.

\section{Examples from particle physics}
\label{sec:examples}

In this section I discuss some examples from particle physics.
The reader not familiar with this field may skip this section.
As before, all examples will be very elementary.

\subsection{Calculation of Born matrix elements}
\label{subsec:born}

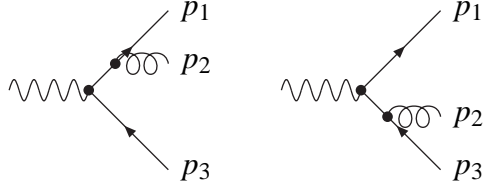
\begin{figure}
\begin{center}
\begin{picture}(100,70)(0,30)
\Photon(10,50)(40,50){4}{4}
\Vertex(40,50){2}
\ArrowLine(40,50)(70,80)
\ArrowLine(70,20)(40,50)
\Vertex(50,60){2}
\Gluon(50,60)(70,60){4}{2}
\Text(75,80)[l]{$p_1$}
\Text(75,60)[l]{$p_2$}
\Text(75,20)[l]{$p_3$}
\end{picture} 
\begin{picture}(100,70)(0,30)
\Photon(10,50)(40,50){4}{4}
\Vertex(40,50){2}
\ArrowLine(40,50)(70,80)
\ArrowLine(70,20)(40,50)
\Vertex(50,40){2}
\Gluon(50,40)(70,40){4}{2}
\Text(75,80)[l]{$p_1$}
\Text(75,40)[l]{$p_2$}
\Text(75,20)[l]{$p_3$}
\end{picture} 
\end{center}
\caption{\label{fig5} The Feynman diagrams contributing to the process
$\gamma^\ast \rightarrow q g \bar{q}$.}
\end{figure}

Here I discuss how to calculate the Born matrix element for
the process $\gamma^\ast \rightarrow q g \bar{q}$.
There are only two contributing Feynman diagrams, which are
shown in fig. \ref{fig5}.
The following script calculates the Born matrix element squared:
\begin{verbatim}
{
// this script calculates the Born matrix element squared for
// photon -> quark gluon antiquark
// trivial coupling and colour prefactors are ignored

ex onehalf = numeric(1,2);
ex D       = numeric(4);

// indices
varidx mu(symbol("mu"),D), nu(symbol("nu"),D);
varidx rho(symbol("rho"),D), sigma(symbol("sigma"),D);

// symbols for fourvectors
symbol p1("p1","p_1"), p2("p2","p_2"), p3("p3","p_3");
symbol p12("p12","p_{12}"), p23("p23","p_{23}");

// symbols for scalarproducts
symbol s12("s12","s_{12}"), s23("s23","s_{23}"), s13("s13","s_{13}");
symbol s123("s123","s_{123}");

// siderelations for simplifications
lst l_mom  = lst( p12==p1+p2, p23==p2+p3 );
lst l_sinv = lst( s13==s123-s12-s23 );

// table for scalarproducts
scalar_products sp;
sp.add(p1,p1,0);
sp.add(p2,p2,0);
sp.add(p3,p3,0);
sp.add(p1,p2,onehalf*s12);
sp.add(p1,p3,onehalf*s13);
sp.add(p2,p3,onehalf*s23);

// polarization sums
ex q_pol_sum      = dirac_slash(p1,D,1);
ex qbar_pol_sum   = dirac_slash(p3,D,1);
ex gluon_pol_sum  = -lorentz_g(rho,sigma);
ex photon_pol_sum = -lorentz_g(mu,nu);

// Feynman diagrams
ex amplitude = (-I) * dirac_gamma(rho.toggle_variance(),1)
     * I / s12 * dirac_slash(p12,D,1)
     * (-I) * dirac_gamma(mu.toggle_variance(),1)
   + 
     (-I) * dirac_gamma(mu.toggle_variance(),1)
     * I / s23 * dirac_slash(-p23,D,1)
     * (-I) * dirac_gamma(rho.toggle_variance(),1);

ex amplitude_conj = I * dirac_gamma(nu.toggle_variance(),1)
     * (-I) / s12 * dirac_slash(p12,D,1)
     * I * dirac_gamma(sigma.toggle_variance(),1)
   +
     I * dirac_gamma(sigma.toggle_variance(),1)
     * (-I) / s23 * dirac_slash(-p23,D,1)
     * I * dirac_gamma(nu.toggle_variance(),1);

// matrix element squared
ex M3_raw = amplitude * qbar_pol_sum * amplitude_conj * q_pol_sum
           * gluon_pol_sum * photon_pol_sum;

// substitute and expand
ex M3_temp = M3_raw.subs(l_mom).expand(expand_options::expand_indexed);

// take trace and simplify
ex M3 = dirac_trace(M3_temp,1).simplify_indexed(sp).subs(l_sinv).expand();
}
\end{verbatim}
Running this script in gTybalt will yield
\bq
M_3 & = &
  16 \frac{s_{123}^2}{s_{12}s_{23}}
 -16 \frac{s_{123}}{s_{12}}
 + 8 \frac{s_{23}}{s_{12}}
 -16 \frac{s_{123}}{s_{23}}
 + 8 \frac{s_{12}}{s_{23}},
\eq
which is the correct result.
The advantage of an interactive program in the debugging phase is, that
one has access to intermediate variables (like ``\v/M3_raw/'' or ``\v/M3_temp/''),
which can be printed in a human-readable form.
Once debugged, the file (with minor modifications)
can be compiled and linked against the GiNaC-library.
It can then be executed without invoking gTybalt.

\subsection{Integration over phase space}
\label{subsec:phasespace}

The result from the previous subsection can be writen as
\bq
M_3 & = & F\left(\frac{s_{12}}{s_{123}}, \frac{s_{13}}{s_{13}+s_{23}}\right)
 + F\left(\frac{s_{23}}{s_{123}}, \frac{s_{13}}{s_{12}+s_{13}}\right),
\eq
where
\bq
F(y,z) & = & \frac{8}{y} \left[
   \frac{2}{1-z(1-y)} -2 + (1-y)(1-z)
 \right].
\eq
Integrated over the appropriate phase space, this expression gives
a higher order contribution to the process $\gamma^\ast \rightarrow q \bar{q}$.
The appropriate phase space measure in terms of the $y$ and $z$
variables reads
\bq
d\phi_3 & = & \frac{s_{123}}{128 \pi^3}
 \int\limits_0^1 dy \; (1-y) \int\limits_0^1 dz.
\eq
This integral diverges for $y=0$ and a standard procedure 
(``phase space slicing'') splits the integral into two regions,
the interval $[0,y_{min}]$ and the intervall $[y_{min},1]$.
In the first region the integral is treated analytically in $D$
dimensions and combined with virtual loop corrections such that
the infrared divergences cancel in the sum.
This will not be discussed here further.
The second region gives a finite contribution, which depends
on the parameter $y_{min}$ and is treated numerically.
One is therefore interested in evaluating integrals of the following type:
\bq
\int\limits_{y_{min}}^1 dy \; (1-y) \int\limits_0^1 dz \; F(y,z)
\eq
The following script ``\v/phasespace.C/''
will perform this integration numerically:
\begin{verbatim}
{
symbol y("y"), z("z");
double ymin = 0.01;

ex integrand = (1-y) * 8/y * (2/(1-z*(1-y))-2+(1-y)*(1-z));
ex result = intnum(integrand,y,z,ymin,1,0,1);
}
\end{verbatim}
One obtains:
\begin{verbatim}
gTybalt> .x phasespace.C
gTybalt> print(gTybalt_int_res);
124.322
gTybalt> print(gTybalt_int_err);
0.00168332
\end{verbatim}
For this simple example the exact result can be calculated, which 
for $y_{min}=0.01$ is 
approximately $124.318$.

\subsection{Loop integrals}
\label{subsec:loop}

Loop integrals occur in higher orders in perturbation theory.
An example for a loop integral is given by the following three-point function
\bq
I & = & 
 \frac{\Gamma(1-2\eps)}{\Gamma(1+\eps)\Gamma(1-\eps)^2}
 \left( - s_{123} \right)^{\nu_{123}-m+\eps}
 \int \frac{d^Dk_1}{i \pi^{D/2}}
 \frac{1}{(-k_1^2)^{\nu_1}}
 \frac{1}{(-k_2^2)^{\nu_2}}
 \frac{1}{(-k_3^2)^{\nu_3}},
\eq
where $k_1=k_2+p_1+p_2$, $k_2=k_3+p_3$, $D=2m-2\eps$, $s_{123}=(p_1+p_2+p_3)^2$,
$\nu_{123}=\nu_1+\nu_2+\nu_3$ and $p_1^2=p_2^2=p_3^2=0$.
\begin{figure}
\begin{center}
\begin{picture}(100,70)(0,30)
\Vertex(20,50){2}
\Vertex(70,80){2}
\Vertex(70,20){2}
\Line(20,50)(70,80)
\Line(70,80)(70,20)
\Line(70,20)(20,50)
\Line(10,50)(20,50)
\Line(10,55)(20,50)
\Line(10,45)(20,50)
\Line(70,80)(80,85)
\Line(70,80)(80,75)
\Line(70,20)(80,20)
\Text(85,80)[l]{$p_1+p_2$}
\Text(85,20)[l]{$p_3$}
\end{picture} 
\end{center}
\caption{\label{fig6} A one-loop integral with two external off-shell legs.}
\end{figure}
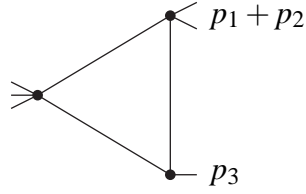    
A diagram for this integral is shown in fig. \ref{fig6}.
It is not too complicated to show that this integral evaluates to the 
following hypergeometric function
\bq
I & = &
 \frac{\Gamma(1-2\eps)}{\Gamma(1+\eps)\Gamma(1-\eps)^2}
 \frac{1}{\Gamma(\nu_1)\Gamma(\nu_2)}
 \frac{\Gamma(m-\eps-\nu_1)\Gamma(m-\eps-\nu_{23})}{\Gamma(2m-2\eps-\nu_{123})}
 \nonumber \\
 & & \times
 \sum\limits_{n=0}^\infty
 \frac{\Gamma(n+\nu_2)\Gamma(n-m+\eps+\nu_{123})}
      {\Gamma(n+1)\Gamma(n+\nu_{23})}
 \left(1-x\right)^n,
\eq
where $x=s_{12}/s_{123}$.
The task is now to expand this expression for specific integer
values of $\nu_1$, $\nu_2$, $\nu_3$ and $m$ in the small parameter $\eps$.
With the help of the nestedsums library this can be done as follows:
\begin{verbatim}
{
symbol x("x"), eps("epsilon");
ex nu1 = 1;
ex nu2 = 1;
ex nu3 = 1;
ex m   = 2;

ex nu23  = nu2+nu3;
ex nu123 = nu1+nu2+nu3;

transcendental_fct_type_A F21(1-x,lst(nu2,nu123-m+eps),lst(nu23),
        lst(1-2*eps,m-eps-nu1,m-eps-nu23),
        lst(1+eps,1-eps,1-eps,nu1,nu2,2*m-2*eps-nu123));
ex result = F21.set_expansion(eps,1);
}
\end{verbatim} 
This script calculates the expansion for $\nu_1=\nu_2=\nu_3=1$ and $m=2$
to order $\eps^0$ and one obtains
\bq
\frac{\mbox{Li}_1(1-x)}{\eps(-1+x)}
+ \frac{S_{0,2}(1-x)}{-1+x}
+ {\cal O}(\eps).
\eq
Expressing the Nielsen polylog $S_{0,2}(1-x)$ in terms of standard
logarithms this is equal to
\bq
\frac{\ln x}{\eps(1-x)} - \frac{\ln^2 x}{2(1-x)}
+ {\cal O}(\eps)
\eq
and is the correct result.

\section{Design of the program}
\label{sec:design}

This section gives some technical details on the design of the program
and serves as a guide to the source code.
The reader who is primarily interested in using the program just as an
application may skip this section in a first reading.
After a general overview of the system I discuss two technical points
concerning threads and dynamic loading, where a few explanations 
might be useful to understand the source code.

\subsection{Structural overview}
\label{sec:struc}

\begin{figure}
\begin{center}
\begin{picture}(400,250)(0,0)
\Text(200,250)[c]{TeXmacs}
\Text(200,200)[c]{gTybalt-bin}
\Text(200,150)[c]{CINT}
\Text(200,100)[c]{gTybalt-dictionary}
\Text(50,50)[c]{GiNaC}
\Text(150,50)[c]{Root}
\Text(250,50)[c]{gTybalt-lib}
\Text(360,50)[c]{Nestedsums}
\Text(50,0)[c]{CLN}
\Text(170,0)[c]{Eqascii}
\Text(250,0)[c]{GSL}
\Text(330,0)[c]{NTL}
\LongArrow(200,235)(200,215)
\LongArrow(200,215)(200,235)
\LongArrow(200,185)(200,165)
\LongArrow(200,165)(200,185)
\LongArrow(200,135)(200,115)
\LongArrow(200,115)(200,135)
\LongArrow(170,85)(65,65)
\LongArrow(65,65)(170,85)
\LongArrow(190,85)(155,65)
\LongArrow(155,65)(190,85)
\LongArrow(210,85)(245,65)
\LongArrow(245,65)(210,85)
\LongArrow(230,85)(335,65)
\LongArrow(335,65)(230,85)
\LongArrow(50,15)(50,35)
\LongArrow(50,35)(50,15)
\LongArrow(250,15)(250,35)
\LongArrow(250,35)(250,15)
\LongArrow(180,15)(230,35)
\LongArrow(230,35)(180,15)
\LongArrow(320,15)(270,35)
\LongArrow(270,35)(320,15)
\end{picture}
\end{center}
\caption{\label{fig7} Structural overview for gTybalt.}
\end{figure}
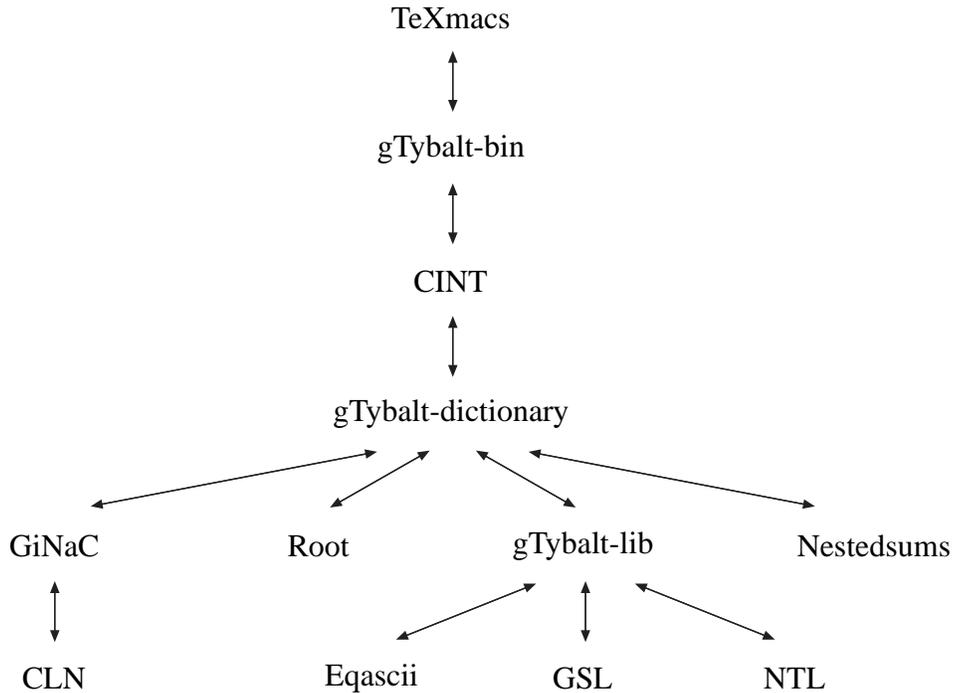

A structural overview for gTybalt is shown in fig. \ref{fig7}.
gTybalt consists of three parts, labelled
\v/gTybalt-bin/, \v/gTybalt-dictionary/ and \v/gTybalt-lib/,
which ensure communication between the different modules
on which gTybalt is based.
The first part, \v/gTybalt-bin/ is either called from TeXmacs
(in TeXmacs-mode) or directly from the shell (when issuing the
command ``\v/gtybalt/'' from a shell) and implements an event
loop.
This program reads input from the keyboard, sends the commands
to the C++ interpreter CINT for execution and directs the output
either to TeXmacs or to a text window.
\\
The program CINT interprets the commands.  
For this purpose it uses a library called 
\\
\v/gTybalt-dictionary/, which can be thought of as a look-up
table where to find the actual implementations of the encountered
function calls.
The source code for this library is generated automatically during
the build phase of gTybalt.
A file ``\v/LinkDef.h/'' specifies which functions and classes 
to include into this library.
The library is then generated from the header files for these
functions and classes.
The CINT interpreter is not $100\%$ standard C++ compatible
and there are certain constructs, which cannot be processed by
CINT.
Therefore the header files for the GiNaC-library are first copied
to a temporary directory and then processed by a perl script, which
comments out any parts which cannot be fed into CINT.
\\
Finally, the library \v/gTybalt-lib/ is an ordinary library, 
defining gTybalt-specific functions like \v/print/, \v/factorpoly/
or \v/intnum/.
It depends in turn on other libraries, like Eqascii, GSL or NTL,
which are however not visible in the interactive interface.

\subsection{Threads and the plotting routine}
\label{sec:threads}

When plotting a function, it is desirable to have the window
with the plot appearing on the screen, but at the same time
still be able to work in the main window of gTybalt.
Since there are now two possible actions which the user
can take (e.g. typing new commands in the main window
and modifying the plot inside the window with the plot)
this is implemented using different threads.
When starting gTybalt, the program will create a separate thread,
which waits on a condition that a function should be plotted.
When a plotting command is issued, CINT invokes a function,
which just prepares some variables for the plot, signals
that there is something to be plotted and then returns.
Therefore after the return of this function
the user can issue new commands in the main window
of gTybalt.
The thread waiting on the condition for plotting a function will
wake up, plot the function and provide an event handler
for events concerning the window with the plot.
Therefore the user can now take actions in both the main window 
for gTybalt and the window with the plot.
Once the window with the plot is cleared (by choosing from the menubar
of the plot ``File $\rightarrow$ Quit ROOT'') the thread for plots
will fall into sleep again and wait till another plotting command
is issued.
Thread safety is guarenteed by copying the relevant expressions
for the function to be plotted to global variables and by the
reference counting mechanism of GiNaC:
The expression to be plotted will be pointed at by at least one
(global) variable, therefore it will not be modified.
While a plot is displayed on the screen, any command to plot
another function will be ignored.
The user must first clear the window with the plot.

\subsection{Dynamic loading of modules and numerical integration}
\label{sec:dlopen}

The default behaviour for numerical evaluation of a function uses
the arbitrary precision arithmetic provided by the CLN library.
For Monte Carlo integration, where a function needs to be evaluated
many times, this is quite slow and therefore inefficient.
It is also not needed, since statistical errors and not rounding
errors tend to dominate the error of the final result.
Therefore a different approach has been implemented for the numerical
Monte Carlo integration:
The function to be integrated is first written as C code to a file,
this file is then compiled with a standard C compiler and the resulting
executable is loaded dynamically (e.g. as a ``plug-in'') into the memory
space of gTybalt and the Monte Carlo integration routine uses this
compiled C function for the evaluations.

\section{Summary}
\label{sec:conclusions}

In this article I discussed the free computer algebra system ``gTybalt''.
It has a modular structure and is based on other, freely available
packages.
gTybalt offers the possibility of interactive symbolic calculations
within the C++ programming language.
Mathematical formulae are visualized using TeX fonts.
It can be extended easily by adding new libraries to it.

\subsection*{Acknowledgements}
\label{sec:acknowledgements}

The program described here relies on other programs.
I would like to thank the authors of these packages:
Ch. Bauer,
A. Frink,
R. Kreckel,
J. van der Hoeven,
M. Goto,
R. Brun,
F. Rademakers,
B. Haible,
P. Borys,
V. Shoup,
M. Galassi,
J. Davies,
J. Theiler,
B. Gough,
G. Jungman,
M. Booth
and F. Rossi.

\begin{appendix}

\section{Installation}
\label{sec:installation}

\subsection{Prerequisities}
\label{subsec:Prerequisities}

gTybalt uses TeXmacs, Root, GiNaC and the GNU scientific library. 
These packages are needed before one starts to build gTybalt.
TeXmacs depends on LaTeX and Guile Scheme, GiNaC on the CLN library.
One therefore needs also these packages.
The C++/C interpreter CINT is not needed as a separate package,
since it is already included in Root.

Building gTybalt requires a 
ANSI-compliant C++-compiler. 
It is strongly recommended to use gcc. 
The version number of the gcc compiler can be obtained by
\begin{verbatim}
$ gcc -v
\end{verbatim}
It is recommended to use version gcc 2.95.3. 
There have been problems reported with versions 2.96 and 3.x.

gTybalt \gversion works with
\begin{verbatim}
 gcc 2.95.3,
 GiNaC 1.0.14,
 Root 3.03.09,
 TeXmacs 1.0.1,
 GSL 1.2,
 CLN 1.1.5,
 NTL 5.0c,
 nestedsums 1.1.1.
\end{verbatim}
gTybalt has been tested with these versions and you are advised to use
EXACTLY those.
\\
You can follow the steps below to build all relevant packages in the
right order.
\\
You need LaTeX to be installed. You can check whether the LaTeX binaries
exist in your path by
\begin{verbatim}
$ which latex
\end{verbatim}
Usually LaTeX is already included in a standard Linux distribution.
\\
You further need Guile Scheme to be installed. You can check by
\begin{verbatim}
$ which guile
\end{verbatim}
If Guile is not installed, you can download it from 
\begin{verbatim}
 http://www.gnu.org/software/guile
\end{verbatim}
and follow the installation instructions.
\\
Installation of TeXmacs: Having LaTeX and Guile Scheme installed, 
you can build and install TeXmacs. TeXmacs can be obtained from 
\begin{verbatim}
 http://www.texmacs.org
\end{verbatim}
Follow the installations instructions there.
\\
Installation of Root: The Root package can be obtained from
\begin{verbatim}
 ftp://root.cern.ch/root
\end{verbatim}
Follow the installation instructions there.
Note that the configuration command has a slightly different syntax
as compared to the standard autotools configuration script.
Further note that the Root package includes also the C++/C interpreter
CINT.
\\
Installation of the CLN library: 
GiNaC requires the CLN library, available from either
one of the following FTP-sites:
\begin{verbatim}
 ftp://ftp.santafe.edu/pub/gnu/
 ftp://ftp.ilog.fr/pub/Users/haible/gnu/ 
 ftp://ftpthep.physik.uni-mainz.de/pub/gnu/
\end{verbatim}
Follow the installation instructions in the manual.
\\
Installation of GiNaC: 
You find GiNaC at
\begin{verbatim}
 http://www.ginac.de
\end{verbatim}
Unpack the GiNaC source code and follow the installation instructions 
of GiNaC.
There is no need to build the package ginac-cint, gTybalt uses the
CINT implementation of Root and builds the necessary dictionary
later.
\\
Installation of the GNU scientific library:
You find the GNU scientific library at
\begin{verbatim}
 http://sources.redhat.com/gsl
\end{verbatim}
Unpack the source code and follow the installation instructions 
of the GNU scientific library.
\\
Optionally gTybalt can be compiled with support for factorization of polynomials.
In this case, the NTL library has to be installed.
The NTL library is available from
\begin{verbatim}
 http://www.shoup.net/ntl
\end{verbatim}
Configure the NTL library with the option "\v/NTL_STD_CXX=on/". This puts the classes
of the NTL library into a namespace "\v/NTL/". 
This option is needed by gTybalt.
\\
Optionally gTybalt can be compiled with support for 
the expansion of transcendental functions.
In this case, the nestedsums library has to be installed.
The nestedsums library is available from
\begin{verbatim}
 http://www.fis.unipr.it/~stefanw/nestedsums
\end{verbatim}
Follow the installations instructions there.

\subsection{Building gTybalt}
\label{subsec:Building}

As with any autoconfiguring GNU software, installation proceeds through
the following steps:
\begin{verbatim}
$ ./configure
$ make
[become root if necessary]
$ make install
\end{verbatim}
The "configure" script can be given a number of options to enable and
disable various features. For a complete list, type:
\begin{verbatim}
$ ./configure --help
\end{verbatim}
A few of the more important ones:
\begin{itemize}
\item
\v/--prefix=PREFIX/: install architecture-independent files in \v/PREFIX/
[defaults to 
\\
{\tt /usr/local}]
\item
\v/--exec-prefix=EPREFIX/: install architecture-dependent files in \v/EPREFIX/ [defaults to the value given to \v/--prefix/]
\item
\v/--with-nestedsums/: If the nestedsums library is installed, one may give this option to enable support for the
expansion of transcendental functions.
\item
\v/--with-ntl/: If the NTL library is installed, one may give this option to enable support for the
factorization of polynomials.
\item
\v/--with-ntl-prefix=PREFIX/: If the NTL library is installed in an unusual location, the value
of \v/PREFIX/ informs the configure script where to find the NTL library.
\item
\v/--with-gmp-prefix=PREFIX/: The NTL library can be compiled with the 
GMP library (GNU Multiple Precision Arithmetic Library) \cite{GMP}. 
If the GMP library
is installed in an unusual location, the value
of \v/PREFIX/ informs the configure script where to find the GMP library.
\end{itemize}
If you install the executables and the libraries in non-standard directories,
you have to set the \v/PATH/ and the \v/LD_LIBRARY_PATH/ variables correspondingly.

\subsection{Example of a complete installation}
\label{subsec:complete}

Here is an example of a complete installation. We assume that LaTeX is already installed and that a suitable compiler
exists.
We further assume that we do not have root privileges, therefore we have to install the programs in an unusual
directory, which we take to be
\\
{\tt /home/username/local}.
We further decide to build only shared libraries (where this is possible) and we would like to set up 
gTybalt with support for factorization and for the expansion of transcendental functions. 
This requires to install the NTL library and the nestedsums library.
We decide to build
the NTL library in conjunction with the GMP (GNU Multi-Precision library) for enhanced performance.
Furthermore we assume that we have unpacked all relevant source packages into separate directories.
The commands to build a specific package are entered from the top-level directory of the specific
package.
For concreteness we assume that we have a Linux PC with a Bourne compatible shell.
\\
We start with Guile:
\begin{verbatim}
$ ./configure --prefix=/home/username/local
$ make
$ make install
\end{verbatim}
We then set the environment variables (probably in a file like \v/.bash_profile/):
\begin{verbatim}
export GUILE_LOAD_PATH=/home/username/local
export PATH=/home/username/local/bin:$PATH
export LD_LIBRARY_PATH=/home/username/local/lib:$LD_LIBRARY_PATH
\end{verbatim}
Building TeXmacs is done in a similar way
\begin{verbatim}
$ ./configure --prefix=/home/username/local
$ make
$ make install
\end{verbatim}
In the next step we build the Root package (note the slightly different syntax):
\begin{verbatim}
$ ./configure linux --prefix=/home/username/local 
  --etcdir=/home/username/local/etc
$ gmake
$ gmake install
\end{verbatim}
and we set the variable {\tt ROOTSYS} (probably in a file like \v/.bash_profile/):
\begin{verbatim}
export ROOTSYS=/home/username/local
\end{verbatim}
The GMP library is build as follows:
\begin{verbatim}
$ ./configure --prefix=/home/username/local
$ make
$ make install
\end{verbatim}
To build the NTL library we issue the commands:
\begin{verbatim}
$ cd src
$ ./configure PREFIX=/home/username/local NTL_STD_CXX=on 
  NTL_GMP_LIP=on GMP_PREFIX=/home/username/local
$ make
$ make install
\end{verbatim}
Note the slight modified form how the arguments are passed to the configure script.
\\
To build the CLN-library:
\begin{verbatim}
$ ./configure --prefix=/home/username/local --enable-static=no
$ make
$ make install
\end{verbatim}
To build GiNaC:
\begin{verbatim}
$ ./configure --prefix=/home/username/local --enable-static=no
$ make
$ make install
\end{verbatim}
To build the GNU scientific library:
\begin{verbatim}
$ ./configure --prefix=/home/username/local --enable-static=no
$ make
$ make install
\end{verbatim}
To build the nestedsums library:
\begin{verbatim}
$ ./configure --prefix=/home/username/local --enable-static=no
$ make
$ make install
\end{verbatim}
And finally we build gTybalt:
\begin{verbatim}
$ ./configure --prefix=/home/username/local --enable-static=no 
  --with-nestedsums --with-ntl 
  --with-ntl-prefix=/home/username/local 
  --with-gmp-prefix=/home/username/local
$ make
$ make install
\end{verbatim}
This completes the installation procedure.

\end{appendix}


\end{document}